\documentclass[prd,reprint,twocolumn,superscriptaddress,letterpaper,nofootinbib,
               nopreprintnumbers,showpacs]{revtex4-1}

\usepackage{amsmath}
\usepackage{microtype}
\usepackage{siunitx}
\usepackage{txfonts}
\usepackage{verbatim}
\usepackage{dcolumn}

\usepackage[usenames, dvipsnames, svgnames, table]{xcolor}
\usepackage[colorlinks=true, citecolor=RoyalBlue,
            linkcolor=BrickRed, urlcolor=ForestGreen]{hyperref}
\usepackage{graphicx}

\usepackage{float}
\usepackage{lineno}
\usepackage{multirow}

\usepackage{acronym}
\usepackage{ifthen}
\usepackage{svn-multi}

\svnid{$Id: P1500248_EarlyaLIGOCalUncertainty.tex 3429 2016-10-07 20:52:12Z jeffrey.kissel@LIGO.ORG $}

\newcommand{\hl}[1]{{#1}}

\newcommand{\rmi}{\mathrm{i}}

\newcommand*\patchAmsMathEnvironmentForLineno[1]{%
  \expandafter\let\csname old#1\expandafter\endcsname\csname #1\endcsname
  \expandafter\let\csname oldend#1\expandafter\endcsname\csname end#1\endcsname
  \renewenvironment{#1}%
     {\linenomath\csname old#1\endcsname}%
     {\csname oldend#1\endcsname\endlinenomath}}%
\newcommand*\patchBothAmsMathEnvironmentsForLineno[1]{%
  \patchAmsMathEnvironmentForLineno{#1}%
  \patchAmsMathEnvironmentForLineno{#1*}}%
\AtBeginDocument{%
\patchBothAmsMathEnvironmentsForLineno{equation}%
\patchBothAmsMathEnvironmentsForLineno{align}%
\patchBothAmsMathEnvironmentsForLineno{flalign}%
\patchBothAmsMathEnvironmentsForLineno{alignat}%
\patchBothAmsMathEnvironmentsForLineno{gather}%
\patchBothAmsMathEnvironmentsForLineno{multline}%
}

\newcommand\OBSSTART{{September 12}}
\newcommand\OBSEND{{October 20}}
\newcommand\OBSEVENTFULLDATE{{September 14, 2015 09:50:45 UTC}}
\newcommand{\OBSDAYS}{{\ensuremath{{16}~\mathrm{days}}}}

\begin{document}

\title{Calibration of the Advanced LIGO detectors for the discovery of the binary black-hole merger GW150914}
%
%

\author{%
B.~P.~Abbott,$^{1}$  
R.~Abbott,$^{1}$  
T.~D.~Abbott,$^{2}$  
M.~R.~Abernathy,$^{1}$  
K.~Ackley,$^{3}$  
C.~Adams,$^{4}$  
P.~Addesso,$^{5}$  
R.~X.~Adhikari,$^{1}$  
V.~B.~Adya,$^{6}$  
C.~Affeldt,$^{6}$
% ...
\\
\medskip
{{}$^{\dag}$Deceased, May 2015. {}$^{\ddag}$Deceased, March 2015. }%
}\noaffiliation
\collaboration{LIGO Scientific Collaboration}
\email[Corresponding Author:~]{lsc-spokesperson@ligo.org}
\affiliation {LIGO, California Institute of Technology, Pasadena, CA 91125, USA }
\affiliation {Louisiana State University, Baton Rouge, LA 70803, USA }
\affiliation {University of Florida, Gainesville, FL 32611, USA }
\affiliation {LIGO Livingston Observatory, Livingston, LA 70754, USA }
\affiliation {University of Sannio at Benevento, I-82100 Benevento, Italy and INFN, Sezione di Napoli, I-80100 Napoli, Italy }
\affiliation {Albert-Einstein-Institut, Max-Planck-Institut f\"ur Gravi\-ta\-tions\-physik, D-30167 Hannover, Germany }
\affiliation {LIGO, Massachusetts Institute of Technology, Cambridge, MA 02139, USA }
\affiliation {Instituto Nacional de Pesquisas Espaciais, 12227-010 S\~{a}o Jos\'{e} dos Campos, SP, Brazil }
\affiliation {Inter-University Centre for Astronomy and Astrophysics, Pune 411007, India }
\affiliation {International Centre for Theoretical Sciences, Tata Institute of Fundamental Research, Bangalore 560012, India }
\affiliation {University of Wisconsin-Milwaukee, Milwaukee, WI 53201, USA }
\affiliation {Leibniz Universit\"at Hannover, D-30167 Hannover, Germany }
\affiliation {Australian National University, Canberra, Australian Capital Territory 0200, Australia }
\affiliation {The University of Mississippi, University, MS 38677, USA }
\affiliation {California State University Fullerton, Fullerton, CA 92831, USA }
\affiliation {Chennai Mathematical Institute, Chennai, India }
\affiliation {University of Southampton, Southampton SO17 1BJ, United Kingdom }
\affiliation {Universit\"at Hamburg, D-22761 Hamburg, Germany }
\affiliation {Albert-Einstein-Institut, Max-Planck-Institut f\"ur Gravitations\-physik, D-14476 Potsdam-Golm, Germany }
\affiliation {Montana State University, Bozeman, MT 59717, USA }
\affiliation {Syracuse University, Syracuse, NY 13244, USA }
\affiliation {SUPA, University of Glasgow, Glasgow G12 8QQ, United Kingdom }
\affiliation {LIGO Hanford Observatory, Richland, WA 99352, USA }
\affiliation {Columbia University, New York, NY 10027, USA }
\affiliation {Stanford University, Stanford, CA 94305, USA }
\affiliation {University of Birmingham, Birmingham B15 2TT, United Kingdom }
\affiliation {RRCAT, Indore MP 452013, India }
\affiliation {Faculty of Physics, Lomonosov Moscow State University, Moscow 119991, Russia }
\affiliation {SUPA, University of the West of Scotland, Paisley PA1 2BE, United Kingdom }
\affiliation {University of Western Australia, Crawley, Western Australia 6009, Australia }
\affiliation {MTA E\"otv\"os University, ``Lendulet'' Astrophysics Research Group, Budapest 1117, Hungary }
\affiliation {Washington State University, Pullman, WA 99164, USA }
\affiliation {University of Oregon, Eugene, OR 97403, USA }
\affiliation {University of Maryland, College Park, MD 20742, USA }
\affiliation {Center for Relativistic Astrophysics and School of Physics, Georgia Institute of Technology, Atlanta, GA 30332, USA }
\affiliation {Universitat de les Illes Balears, IAC3---IEEC, E-07122 Palma de Mallorca, Spain }
\affiliation {NASA/Goddard Space Flight Center, Greenbelt, MD 20771, USA }
\affiliation {Canadian Institute for Theoretical Astrophysics, University of Toronto, Toronto, Ontario M5S 3H8, Canada }
\affiliation {Tsinghua University, Beijing 100084, China }
\affiliation {University of Michigan, Ann Arbor, MI 48109, USA }
\affiliation {National Tsing Hua University, Hsinchu City, Taiwan 30013, R.O.C. }
\affiliation {Charles Sturt University, Wagga Wagga, New South Wales 2678, Australia }
\affiliation {University of Chicago, Chicago, IL 60637, USA }
\affiliation {Caltech CaRT, Pasadena, CA 91125, USA }
\affiliation {Korea Institute of Science and Technology Information, Daejeon 305-806, Korea }
\affiliation {Carleton College, Northfield, MN 55057, USA }
\affiliation {University of Brussels, Brussels 1050, Belgium }
\affiliation {Sonoma State University, Rohnert Park, CA 94928, USA }
\affiliation {Texas Tech University, Lubbock, TX 79409, USA }
\affiliation {Cardiff University, Cardiff CF24 3AA, United Kingdom }
\affiliation {University of Minnesota, Minneapolis, MN 55455, USA }
\affiliation {The University of Melbourne, Parkville, Victoria 3010, Australia }
\affiliation {The University of Texas Rio Grande Valley, Brownsville, TX 78520, USA }
\affiliation {The University of Sheffield, Sheffield S10 2TN, United Kingdom }
\affiliation {Montclair State University, Montclair, NJ 07043, USA }
\affiliation {The Pennsylvania State University, University Park, PA 16802, USA }
\affiliation {School of Mathematics, University of Edinburgh, Edinburgh EH9 3FD, United Kingdom }
\affiliation {Indian Institute of Technology, Gandhinagar Ahmedabad Gujarat 382424, India }
\affiliation {Institute for Plasma Research, Bhat, Gandhinagar 382428, India }
\affiliation {University of Szeged, D\'om t\'er 9, Szeged 6720, Hungary }
\affiliation {Embry-Riddle Aeronautical University, Prescott, AZ 86301, USA }
\affiliation {Tata Institute of Fundamental Research, Mumbai 400005, India }
\affiliation {American University, Washington, D.C. 20016, USA }
\affiliation {University of Massachusetts-Amherst, Amherst, MA 01003, USA }
\affiliation {University of Adelaide, Adelaide, South Australia 5005, Australia }
\affiliation {West Virginia University, Morgantown, WV 26506, USA }
\affiliation {SUPA, University of Strathclyde, Glasgow G1 1XQ, United Kingdom }
\affiliation {IISER-TVM, CET Campus, Trivandrum Kerala 695016, India }
\affiliation {Northwestern University, Evanston, IL 60208, USA }
\affiliation {Institute of Applied Physics, Nizhny Novgorod, 603950, Russia }
\affiliation {Pusan National University, Busan 609-735, Korea }
\affiliation {Hanyang University, Seoul 133-791, Korea }
\affiliation {Rochester Institute of Technology, Rochester, NY 14623, USA }
\affiliation {Monash University, Victoria 3800, Australia }
\affiliation {Seoul National University, Seoul 151-742, Korea }
\affiliation {University of Alabama in Huntsville, Huntsville, AL 35899, USA }
\affiliation {Southern University and A\&M College, Baton Rouge, LA 70813, USA }
\affiliation {College of William and Mary, Williamsburg, VA 23187, USA }
\affiliation {Instituto de F\'\i sica Te\'orica, University Estadual Paulista/ICTP South American Institute for Fundamental Research, S\~ao Paulo SP 01140-070, Brazil }
\affiliation {University of Cambridge, Cambridge CB2 1TN, United Kingdom }
\affiliation {IISER-Kolkata, Mohanpur, West Bengal 741252, India }
\affiliation {Rutherford Appleton Laboratory, HSIC, Chilton, Didcot, Oxon OX11 0QX, United Kingdom }
\affiliation {Whitman College, 280 Boyer Ave, Walla Walla, WA 9936, USA }
\affiliation {National Institute for Mathematical Sciences, Daejeon 305-390, Korea }
\affiliation {Hobart and William Smith Colleges, Geneva, NY 14456, USA }
\affiliation {Andrews University, Berrien Springs, MI 49104, USA }
\affiliation {Trinity University, San Antonio, TX 78212, USA }
\affiliation {University of Washington, Seattle, WA 98195, USA }
\affiliation {Kenyon College, Gambier, OH 43022, USA }
\affiliation {Abilene Christian University, Abilene, TX 79699, USA }


\begin{abstract}
    In Advanced LIGO, detection and astrophysical source parameter estimation of the binary black hole merger GW150914 requires a calibrated estimate of the gravitational-wave strain sensed by the detectors.
    Producing an estimate from each detector's differential arm length control loop readout signals requires applying time domain filters, which are designed from a frequency domain model of the detector's gravitational-wave response. 
    The gravitational-wave response model is determined by the detector's opto-mechanical response and the properties of its feedback control system.
    The measurements used to validate the model and characterize its uncertainty are derived primarily from a dedicated photon radiation pressure actuator, with cross-checks provided by optical and radio frequency references.
    We describe how the gravitational-wave readout signal is calibrated into equivalent gravitational-wave-induced strain and how the statistical uncertainties and systematic errors are assessed.
    Detector data collected over 38 calendar days, from {\OBSSTART} to {\OBSEND}, 2015, contain the event GW150914 and approximately {\OBSDAYS} of coincident data used to estimate the event false alarm probability.
    The calibration uncertainty is less than \hl{10\%} in magnitude and \hl{$10^\circ$} in phase across the relevant frequency band, \hl{20~Hz to 1~kHz}.
\end{abstract}


\maketitle

\svnid{$Id: Introduction.tex 3007 2016-06-16 16:47:41Z jeffrey.kissel@LIGO.ORG $}

\section{Introduction}
\label{sec:intro}

\begin{figure}
    \centering
    \includegraphics[width=\columnwidth]{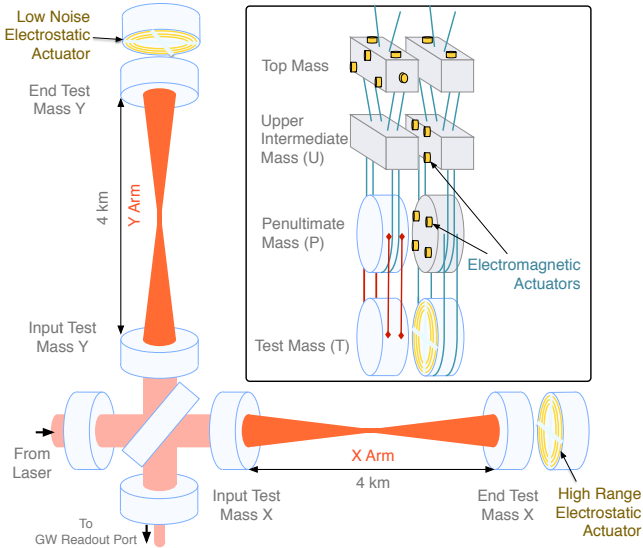}
    \caption{Simplified diagram of an Advanced LIGO interferometer.
    Four highly reflective test masses form two Fabry--P\'{e}rot arm cavities.
    At lower left, a power recycling mirror placed between the laser and the beamsplitter increases the power stored in the arms to 100\,kW.
    A signal recycling mirror, placed between the beamsplitter and the GW readout photodetector, alters the frequency response of the interferometer to differential arm length fluctuations.
    For clarity, only the lowest suspension stage is shown for the optics.
    Inset: one of the dual-chain, quadruple pendulum suspension systems is shown.}
    \label{fig:ifo}
\end{figure}

On \OBSEVENTFULLDATE, the two Advanced LIGO detectors observed a gravitational-wave (GW) signal, GW150914, originating from the merging of two stellar-mass black holes~\cite{GW150914-DETECTION}.
The event was observed in coincident data from the two LIGO detectors between {\OBSSTART} to {\OBSEND}, 2015.
These detectors, H1 located on the Hanford Reservation in Richland, Washington, and L1 located in Livingston Parish, Louisiana, are laser interferometers~\cite{P1400177} that use four mirrors (referred to as test masses) suspended from multi-stage pendulums to form two perpendicular optical cavities (arms) in a modified Michelson configuration, as shown in Fig.~\ref{fig:ifo}.
GW strain causes apparent differential variations of the arm lengths which generate power fluctuations in the interferometer's GW readout port.
These power fluctuations, measured by photodiodes, serve as both the GW readout signal and an error signal for controlling the differential arm length~\cite{P1000009}.

Feedback control of the differential arm length degree of freedom (along with the interferometer's other length and angular degrees of freedom) is required for stable operation of the instrument.
This control is achieved by taking a digitized version of the GW readout signal $d_\text{err}(f)$, applying a set of digital filters to produce a control signal $d_\text{ctrl}(f)$, then sending the control signal to the test mass actuator systems which displace the mirrors.
Without this control system, differential length variations arising from either displacement noise or a passing GW would cause an unsuppressed (free-running) change in differential length, $\Delta L_{\text{free}} = L_x - L_y = h L$, where $L \equiv (L_x + L_y)/2$ is the average length of each detector's arms, with lengths $L_x$ and $L_y$, and $h$ is the sensed strain, $h\equiv\Delta L_{\text{free}}/L$.
In the presence of feedback control, however, this free-running displacement is suppressed to a smaller, residual length change given by $\Delta L_\text{res} = \Delta L_\text{free}(f)/[1+G(f)]$, where $G(f)$ is the open loop transfer function of the differential arm length servo.
Therefore, estimating the equivalent GW strain sensed by the interferometer requires detailed characterization of, and correction for, the effect of this loop. 
The effects of other feedback loops associated with other degrees of freedom are negligible across the relevant frequency band, from \hl{20\,Hz to 1\,kHz}.

\begin{figure}
    \centering
    \includegraphics[width=\columnwidth]{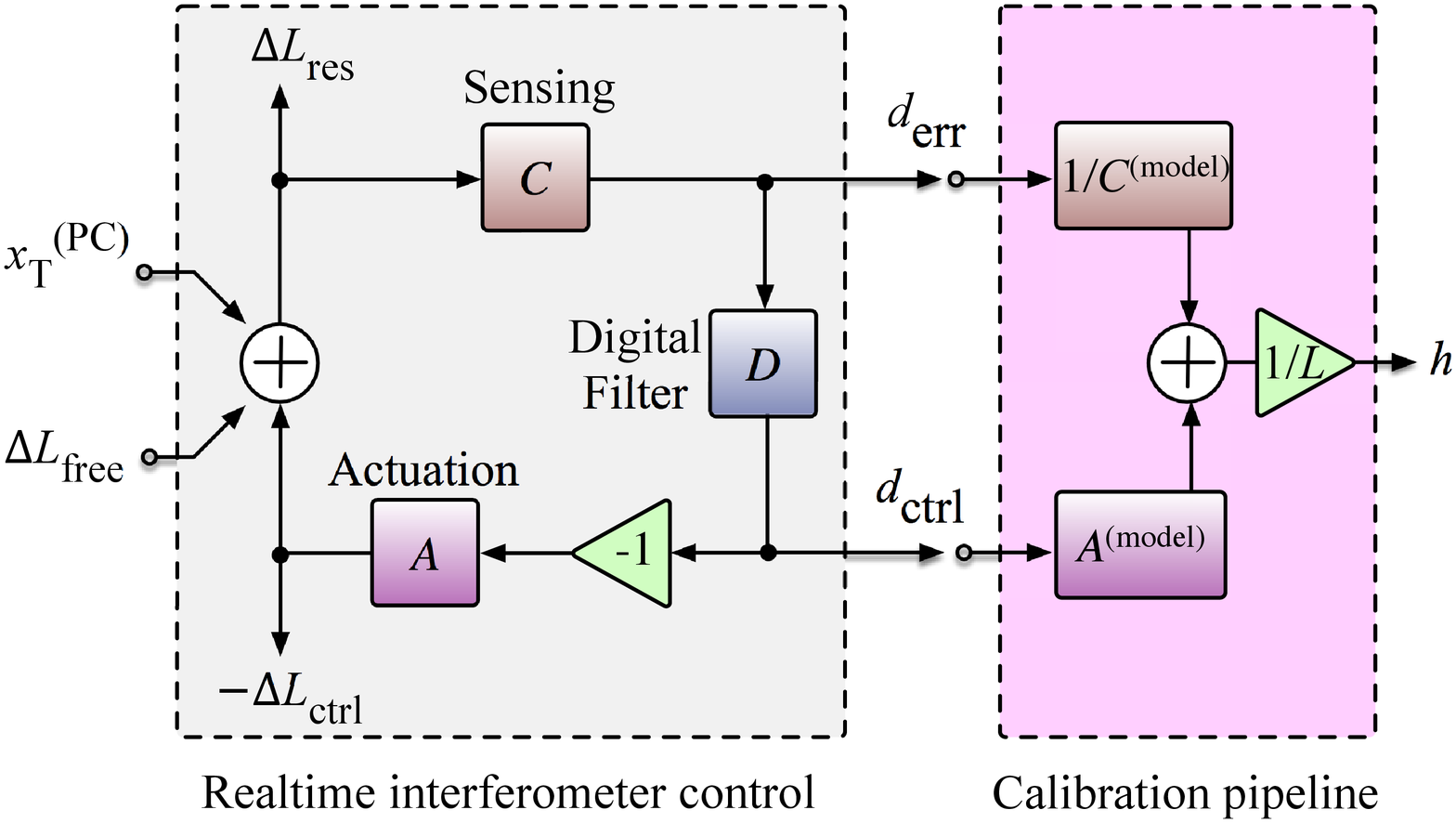}
    \caption{Block diagram of the differential arm length feedback control servo.
        The sensing function, digital filter function, and actuation function combine to form the open loop transfer function $G(f) = A(f)\,D(f)\,C(f)$. The signal $x_\text{T}^\text{(PC)}$ is the modulated displacement of the test masses from the radiation pressure actuator described in Section~\ref{sec:pcal}.
    }
    \label{fig:loopdiagram}
\end{figure}

The differential arm length feedback loop is characterized by a sensing function $C(f)$, a digital filter function $D(f)$, and an actuation function $A(f)$, which together give the open loop transfer function
\begin{equation}
    G(f) = A(f)\,D(f)\,C(f)\,.
    \label{eq:Gdef}
\end{equation}
The sensing function describes how residual arm length displacements propagate to the digitized error signal, $d_\text{err}(f) \equiv C(f)\, \Delta L_\text{res}(f)$; the digital filter function describes how the digital control signal is generated from the digital error signal, $d_\text{ctrl}(f) \equiv D(f)\,d_\text{err}(f)$; and the actuation function describes how the digital control signal produces a differential displacement of the arm lengths, $\Delta L_\text{ctrl} \equiv A(f)\,d_\text{ctrl}(f)$.
These relationships are shown schematically in Fig.~\ref{fig:loopdiagram}.

Either the error signal, the control signal, or a combination of the two can be used estimate the strain sensed by the detector~\cite{LIGOhoft}.
For Advanced LIGO, a combination was chosen that renders the estimate of the detector strain output insensitive to changes in the digital filter function $D$, and makes application of slow corrections to the sensing and actuation functions convenient:
\begin{equation}\label{eq:hoft}
    h(t) = \frac{1}{L} \left[\mathcal{C}^{-1} * d_\text{err}(t) + \mathcal{A} * d_\text{ctrl}(t) \right]\,,
\end{equation}
where $\mathcal{A}$ and $\mathcal{C}^{-1}$ are time domain filters generated from frequency domain models of $A$ and $C$, and $*$ denotes convolution. 

The accuracy and precision of this estimated strain rely on characterizing the sensing and actuation functions of each detector, $C$ and $A$. 
Each function is represented by a model, generated from measurements of control loop parameters, each with associated statistical uncertainty and systematic error. 
Uncertainty in the calibration model parameters directly impacts the uncertainty in the reconstructed detector strain signal.
This uncertainty could limit the signal-to-noise ratios of GW detection statistics, and could dominate uncertainties in estimated astrophysical parameters, e.g., luminosity distance, sky location, component masses, and spin.
Calibration uncertainty is thus crucial for GW searches and parameter estimation.

This paper describes the accuracy and precision of the model parameters and of the estimated detector strain output over the course of the 38 calendar days of observation during which GW150914 was detected. 
Sec.~\ref{sec:model} describes the actuation and sensing function models in terms of their measured parameters.
Sec.~\ref{sec:definitions} defines the treatment of uncertainty and error for each of these parameters.
In Sec.~\ref{sec:pcal}, a description of the radiation pressure actuator is given.
Secs.~\ref{sec:ActuatorCalibration} and \ref{sec:SensorCalibration} discuss the measurements used to determine the static statistical uncertainties and systematic errors in the actuation and sensing function models, respectively, and their results.
Sec.~\ref{sec:dynamic} details the systematic errors in model parameters near the time of the GW150914 event resulting from uncorrected, slow time variations.
Sec.~\ref{sec:total} discusses each detector's strain response function that is used to estimate the overall amplitude and phase uncertainties and systematic errors in the calibrated data stream $h(t)$.
Sec.~\ref{sec:Timing} discusses the inter-site uncertainty in the relative timing of each detector's data stream.
In Sec.~\ref{sec:discussion} the implications of these uncertainties on the detection and astrophysical parameter estimation of GW150914 are summarized.
Finally, in Sec.~\ref{sec:conclusion} we give an outlook on future calibration and its role in GW detection and astrophysical parameter estimation.

\svnid{$Id: ModelDescription.tex 3980 2016-12-19 17:38:00Z evan.goetz@LIGO.ORG $}

\section{Model Description}
\label{sec:model}

We divide the differential arm length feedback loop into two main functions, sensing and actuation. In this section, these functions are described in detail. The interferometer response function is also introduced; it is composed of these functions and the digital control filter function (which is precisely known and carries no uncertainty), and is useful for estimating the overall uncertainty in the estimated strain.

\subsection{Sensing function}

The sensing function $C$ converts residual test mass differential displacement $\Delta L_\text{res}$ to a digitized signal representing the laser power fluctuation at the GW readout port, $d_{\text{err}}$, sampled at a rate of 16~384~Hz. 
It includes the interferometric response converting displacement to laser power fluctuation at the GW readout port, the response of the photodiodes and their analog readout electronics, and effects from the digitization process.

The complete interferometric response is determined by the arm cavity mirror (test mass) reflectivities, the reflectivity of the signal recycling mirror (see Fig.~\ref{fig:ifo}), the length of the arm cavities and the length of the signal recycling cavity~\cite{Mizuno1993,Rakhmanov2008}. The response is approximated by a single-pole low-pass filter with a gain and an additional time delay.

\begin{figure}
    \centering
    \includegraphics[width=0.99\columnwidth]{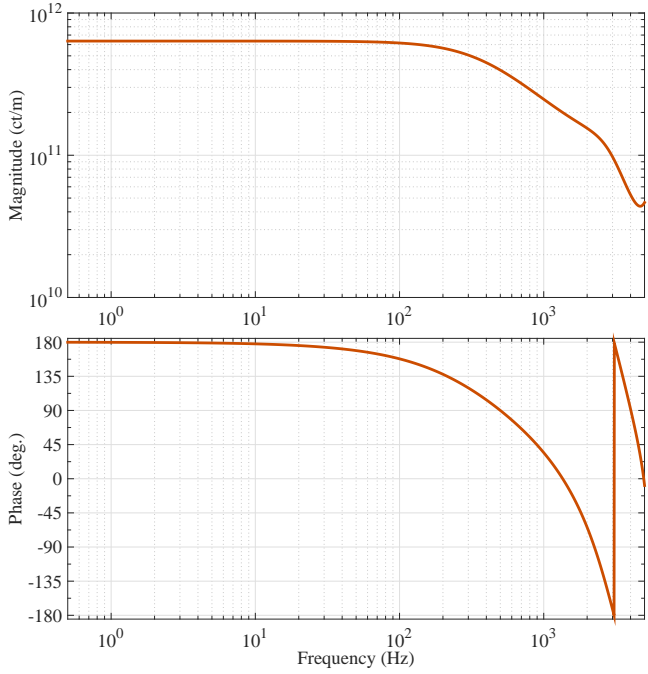}
    \caption{\label{fig:SensingModel}The magnitude and phase of the sensing function model $C(f)$ for the L1 detector. Below 1~kHz the frequency dependence is determined by $f_{\text{C}}$, while above 1~kHz it is determined by the analog-to-digital conversion process.}
\end{figure}

The sensing function is thus given by
\begin{equation}
    C^{(\text{model})}(f) = \frac{\mathcal{K}_C}{1 + \rmi f/f_C} ~C_R(f) \, \exp(-2\piup\rmi f \tau_C)\,,
    \label{eq:Cdef}
\end{equation}
where $\mathcal{K}_C$ is combined gain of the interferometric response and analog-to-digital converter (see Fig.~\ref{fig:SensingModel}). 
It describes, at a reference time, how many digital counts are produced in $d_\text{err}$ in response to differential arm length displacement.
The pole frequency, $f_{\text{C}}$, is the characteristic frequency that describes the attenuation of the interferometer response to high-frequency length perturbations~\cite{Mizuno1993, P1500277}. 
Though each interferometer is designed to have the same pole frequency, the exact value differs as result of discrepant losses in their optical cavities: 341~Hz and 388~Hz for H1 and L1, respectively. 
The time delay $\tau_\text{C}$ includes the light travel time $L/c$ along the length of the arms ($L = 3994.5$~m), computational delay in the digital acquisition system, and the delay introduced to approximate the complete interferometric response as a single pole. 
Finally, the dimensionless quantity $C_R(f)$ accounts for additional frequency dependence of the sensing function above 1~kHz, arising from the properties of the photodiode electronics, as well as analog and digital signal processing filters.

\subsection{Actuation function}

The interferometer differential arm length can be controlled by actuating on the quadruple suspension system for any of the four arm cavity test masses.
Each of these systems consists of four stages, suspended as cascading pendulums~\cite{P020001, P1200056}, which isolate the test mass from residual motion of the supporting active isolation system~\cite{P1200040}. 
Each suspension system also includes an adjacent, nearly-identical, cascaded reaction mass pendulum chain which can be used to independently generate reaction forces on each mass of the test mass pendulum chain.
A diagram of one of these suspension systems is shown in Fig.~\ref{fig:ifo}. 

For each of the three lowest stages of the suspension system---the upper intermediate mass (U), the penultimate mass (P), and the test mass (T)---digital-to-analog converters and associated electronics drive a set of four actuators that work in concert to displace each stage, and consequently the test mass suspended at the bottom.
The digital control signal $d_{\text{ctrl}}$ is distributed to each stage and multiplied by a set of dimensionless digital filters $F_{i}(f)$, where $i =$ U, P, or T, so that the lower stages are used for the highest frequency signal content and the upper stages are used for high-range, low-frequency signal content.

While the differential arm length can be controlled using any combination of the four test mass suspension systems, only one, the Y-arm end test mass, is used to create $\Delta L_{\text{ctrl}}$. Actuating a single test mass affects both the common and the differential arm lengths. The common arm length change is compensated, however, by high-bandwidth ($\sim$14~kHz) feedback to the laser frequency.

\begin{figure}
    \centering
    \includegraphics[width=\columnwidth]{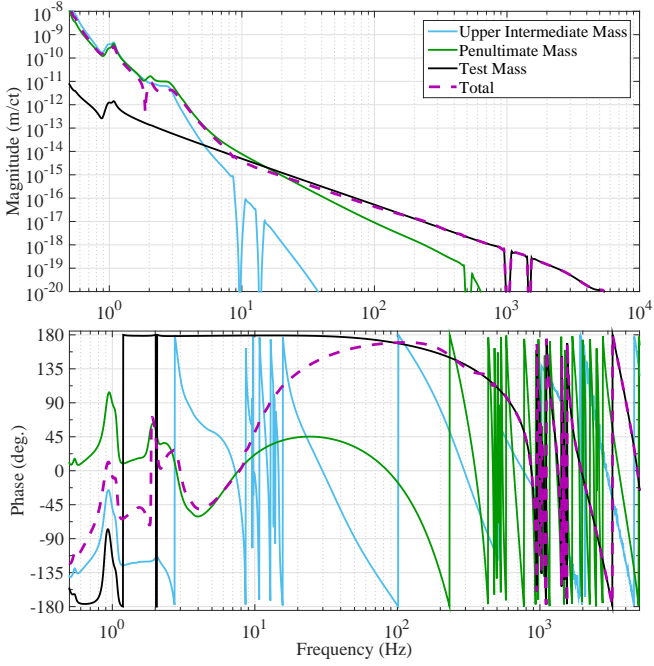}
    \caption{\label{fig:ActuationModel}Overall actuation transfer function $A(f)$ and actuation functions for each suspension stage $F_i(f) \,\mathcal{K}_i \, A_\text{i}(f)$ for the L1 detector. The mechanical response of the pendulums and $F_{i}$ dictate the characteristics of each stage. The strongest actuator, that for the upper intermediate mass, is used below a few Hz. Above $\sim$30\,Hz, only the test mass actuator is used. At certain frequencies (e.g., 10, 14, and 500~Hz), digital notch filters are implemented for high quality factor features of the pendulum responses in order to avoid mechanical instabilities. The H1 actuation function differs slightly in scale, frequency dependence, and digital filter choice.}
\end{figure}

The model of the actuation function $A$ of the suspension system comprises the mechanical dynamics, electronics, and digital filtering, and is written as
\begin{align}
    A^{(\text{model})}(f) &= \Big[ F_\text{T}(f) \,\mathcal{K}_\text{T} \, A_\text{T}(f)
            + F_\text{P}(f) \, \mathcal{K}_\text{P} \, A_\text{P}(f) \nonumber\\
         &\hphantom{=}\hspace{2em} {} + F_\text{U}(f) \, \mathcal{K}_\text{U} \, A_\text{U}(f) \Big] \, \exp(-2\piup\rmi f \tau_{A})\,.
    \label{eq:Adef}
\end{align}
Here $\mathcal{K}_i$ and $A_i(f)$ are the gain and the normalized frequency dependence of the $i$th suspension stage actuator, measured at a reference time, that define the actuation transfer function for each suspension stage; $\tau_{A}$ is the computational delay in the digital-to-analog conversion.
The overall and individual stage actuation functions are plotted as a function of frequency in Fig.~\ref{fig:ActuationModel}.
The gain converts voltage applied at suspension stage $i$ to test mass displacement.
The frequency response is primarily determined by the mechanical dynamics of the suspension, but also includes minor frequency dependent terms from digital-to-analog signal processing, analog electronics, and mechanical interaction with the locally-controlled suspension stage for the top mass (see Fig.~\ref{fig:ifo}).
While opto-mechanical interaction from radiation pressure can affect the actuation function~\cite{buonanno2002signal}, the laser power resonating in the arm cavities during the observation period was low enough that radiation pressure effects can be ignored.
The H1 and L1 suspensions and electronics are identical by design, but there are slight differences, mostly due to the digital filtering for each stage $F_i$, which are precisely known and carry no uncertainty.

\subsection{\label{sec:responseFunc}Response function}
For uncertainty estimation, it is convenient to introduce the response function $R(f)$ that relates the differential arm length servo error signal to strain sensed by the interferometer: $h(f) = (1/L)\,R(f)\,d_\text{err}(f)$.
As shown schematically in Fig.~\ref{fig:loopdiagram}, the response function is given by
\begin{equation}
    R(f) = \frac{1+A(f)\,D(f)\,C(f)}{C(f)} = \frac{1+G(f)}{C(f)}\,.
    \label{eq:Rdef}
\end{equation}
We will use this response function to evaluate the overall accuracy and precision of the calibrated detector strain output.
The actuation function dominates at frequencies below the differential arm length servo unity gain frequency, 40~Hz and 56~Hz for H1 and L1, respectively.
Above the unity gain frequency, the sensing function dominates (see Figs.~\ref{fig:SensingModel} and \ref{fig:ActuationModel}).

\svnid{$Id: UncertaintyDefinitions.tex 2703 2016-02-11 04:39:22Z evan.hall@LIGO.ORG $}

\section{Definitions of Parameter Uncertainty}
\label{sec:definitions}
From Eqs.~\eqref{eq:Cdef} and \eqref{eq:Adef}, we identify the set $Q^{\text{(model)}}$ of parameters shown in Table~\ref{tbl:params} that define the model for each detector's sensing and actuation functions. 
These model parameters have both statistical uncertainty and systematic error.
In this section, we outline how the uncertainty and error for each parameter are treated.
Discussion of how these are propagated to inform the total uncertainty and error in final estimated strain $h(t)$ is left to Section \ref{sec:total}.

\begin{table}
\caption{\label{tbl:params}The set of differential arm length control loop parameters, $Q^{(\text{model})}$ that must be characterized to define the sensing and actuation functions.}
\begin{ruledtabular}
\begin{tabular}{ll}
Parameter & Description \\
\hline
$A_\text{T}(f)$ & Normalized test mass actuation function \\
$A_\text{P}(f)$ & Normalized penultimate mass actuation function \\
$A_\text{U}(f)$ & Normalized upper intermediate mass actuation function \\
$C_\text{R}(f)$ & Residual sensing function frequency dependence \\
$\mathcal{K}_C$ & Sensing function gain \\
$\mathcal{K}_\text{T}$ & Test mass actuation function gain \\
$\mathcal{K}_\text{P}$ & Penultimate mass actuation function gain \\
$\mathcal{K}_\text{U}$ & Upper intermediate mass actuation function gain \\
$f_C$ & Cavity pole frequency \\
$\tau_C$ & Sensing function time delay \\
\end{tabular}
\end{ruledtabular}
\end{table}

Combinations of the model's scalar parameters ($\mathcal{K}_C$, $\mathcal{K}_\text{T}$, $\mathcal{K}_\text{P}$, $\mathcal{K}_\text{U}$, $f_C$, and $\tau_{C}$) and frequency-dependent functions ($A_\text{T}(f)$, $A_\text{P}(f)$, $A_\text{U}(f)$, and $C_\text{R}(f)$) are constrained by a set of directly measurable properties of the detector $Q^{\text{(meas)}}$:
\begin{align}
Q^{(\text{meas})}(f) &= \bigl\{ \mathcal{K}_\text{T}A_\text{T}(f), \nonumber\\
    &\hphantom{=}\hspace{1em} {} \mathcal{K}_\text{P}A_\text{P}(f),  \nonumber\\
    &\hphantom{=}\hspace{1em} {} \mathcal{K}_\text{U}A_\text{U}(f), \nonumber\\
    &\hphantom{=}\hspace{1em} {} \mathcal{K}_\text{C} C_\text{R}(f) / (1 + \rmi f / f_{\text{C}})~\text{exp}(-2 \pi i f \tau_{\text{C}}) \bigr\}\,. 
    \label{eq:measurementdefinitions} 
\end{align} 
The parameters in $Q^{(\text{model})}$ not included in Table~\ref{tbl:params}, $F_{i}(f)$ and $\tau_A$, are part of the digital control system, known with negligible uncertainty, and are thus removed from the measured quantities without consequence.
Each quantity $q_{i}^{(\text{meas})} \in Q^{(\text{meas})}$ is measured using sinusoidal excitations injected at various points in the control loop while the detector is in its lowest noise state.  
The measurements consist of excitations that are injected consecutively at discrete frequencies, $f_{k}$. 
Only measurements made at a reference time $t_0$ are used to determine the corresponding model parameters $q_{i}^{(\text{model})}$, however the measurements are repeated periodically to inform and reduce uncertainty.

The frequency-dependent model parameters $Q^{(\text{model})}$ described in Table~\ref{tbl:params} do not completely describe the frequency-dependent quantities in $Q^{(\text{meas})}$ at the reference time. In addition, the scalar quantities in $Q^{(\text{meas})}$ vary with time after the reference measurement. Both discrepancies are systematic errors, $\delta q_{i}$. Albeit small, they are carried with each parameter $Q^{(\text{model})}$ through to inform the known systematic error in the response function, and quantified in the following fashion.

Any discrepancy between $A_{i}(f)$ and $C_{\text{R}}(f)$ and the measurements exposes poorly modeled properties of the detector, and thus are systematic errors in Eqs.~\eqref{eq:Cdef} and \eqref{eq:Adef}; $\delta q_{i} = q_\text{i}^{(\text{meas})} - q_{i}^{\text{(model)}}$.
We find it convenient to quantify this systematic error in terms of a multiplicative correction factor to  Eqs.~\eqref{eq:Cdef} and \eqref{eq:Adef}, $\zeta_{i}^{\text{(fd)}} \equiv q_{i}^{\text{(meas)}} / q_{i}^{\text{(model)}} \equiv 1 + (\delta q_{i}/q_{i}^{\text{(model)}})$, instead of dealing directly with the systematic error $\delta q_{i}$.
These frequency-dependent discrepancies are confirmed with repeated measurements beyond the reference time.

The scalar parameters, $\mathcal{K}_{i}$ and $f_{\textrm{C}}$, are monitored continuously during data taking to track small, slow temporal variations beyond the reference measurement time $t_{0}$. 
Tracking is achieved using a set of sinusoidal excitations at select frequencies, typically referred to as \emph{calibration lines}. 
The observed time dependence is treated as an additional systematic error, $\delta q_{i}(t)$, also implemented as a correction factor, $\zeta_{i}^{\text{(td)}} \equiv \delta q_{i}(t) / q_{i}^{(\text{model})}$.

In order to quantify the statistical uncertainties in the frequency-dependent parameters in $Q^{\text{(model)}}$, we divide the measurements $Q^{(\text{meas})}$ by the appropriate combination of reference model parameters $q_{i}^{\text{(model)}}$, time-dependent scalar correction factors, $\zeta_{i}^{\text{(td)}}$, and a fit to any frequency-dependent correction factors, $\zeta_{i}^{\text{(fd,fit)}}$ to form a statistical residual,
\begin{equation}
\xi_{i}^{\text{(stat)}} = q_{i}^{\text{(meas)}} / (q_{i}^{\text{(model)}} \zeta_{i}^{\text{(td)}} \zeta_{i}^{\text{(fd,fit)}}) - 1.
\end{equation}
We assume this remaining residual reflects an estimate of the complex, scalar (i.e. frequency \emph{independent}), statistical uncertainty, $\sigma_{q_{i}~q_{j}}$, randomly sampled over the measurement frequency vector $f_{k}$, and may be covariant between parameter $q_{i}^{\text{(meas)}}$ and $q_{j}^{\text{(meas)}}$. Thus, we estimate $\sigma_{q_{i}~q_{j}}$ by computing the standard deviation of the statistical residual, $\xi_{i}^{\text{(stat)}}$, across the frequency band,
\begin{equation}
\sigma_{q_{i}~q_{j}} = \sum_{k=1}^{N} \frac{(\xi_{i}^{\text{(stat)}}(f_{k}) - \overline{\xi_{i}^{\text{(stat)}}})(\xi_{j}^{\text{(stat)}}(f_{k}) - \overline{\xi_{j}^{\text{(stat)}}})}{(N - 1)}
\end{equation}
where $\overline{\xi_{i}^{\text{(stat)}}} = \sum_{k} \xi_{i}^{\text{(stat)}}(f_{k}) / N$ is the mean across the $N$ points in the frequency vector $f_{k}$.

The time-dependent correction factor, $\zeta_{i}^{\text{(td)}}$, has associated statistical uncertainty $\sigma_{\zeta_{i}^{\text{(td)}}}$ that is governed by the signal-to-noise ratio of the continuous excitation.
Only a limited set of lines were used to determine these time-dependent systematic errors, so their estimated statistical uncertainty is also, in general covariant.

In Secs.~\ref{sec:ActuatorCalibration},~\ref{sec:SensorCalibration}, and~\ref{sec:dynamic}, we describe the techniques for measuring $Q^{(\text{meas})}$ at the reference time $t_{0}$, and discuss resulting estimates of statistical uncertainty $\sigma_{q_{i}~q_{j}}$ and systematic error $\delta q_{i}$, via correction factors $\zeta_{i}$, for each detector.
In Sec.~\ref{sec:total}, we describe how the uncertainty and error estimates for these parameters are combined to estimate the overall accuracy and precision of the calibrated detector strain output $h(t)$.

\svnid{$Id: ActuatorCalibration.tex 2731 2016-02-18 19:01:53Z evan.goetz@LIGO.ORG $}

\section{Radiation pressure actuator}
\label{sec:pcal}
The primary method for calibrating the actuation function $A$ and sensing function $C$ is an independent radiation pressure actuator called the \emph{photon calibrator} (PC)~\cite{P1500249}. A similar system was also used for calibration of the initial LIGO detectors~\cite{P080118}.

Each detector is equipped with two photon calibrator systems, one for each end test mass, positioned outside the vacuum enclosure at the ends of the interferometer arms.
For each system, 1047\,nm light from an auxiliary, power-modulated, Nd$^{3+}$:YLF laser is directed into the vacuum envelope and reflects from the front surface of the mirror (test mass).
The reflected light is directed to a power sensor located outside the vacuum enclosure. This sensor is an InGaAs photodetector mounted on an integrating sphere and is calibrated using a standard that is traceable to the National Institute of Standards and Technology (NIST).
Power modulation is accomplished via an acousto-optic modulator that is part of an optical follower servo that ensures that the power modulation follows the requested waveform.
After modulation, the laser beam is divided optically and projected onto the mirror in two diametrically opposed positions. The spots are separated vertically, ${\pm}11.6$\,cm from the center of the optical surface, on the nodal ring of the drumhead elastic body mode, to minimize errors at high-frequency caused by bulk deformation~\cite{Hutchinson,Hild,P080118,daveloza2012controlling}.

The laser power modulation induces a modulated displacement of the test mass that is given by~\cite{P080118}
\begin{equation}
    x_\text{T}^\text{(PC)}(f) = \frac{2P(f)}{c}\,s(f)\,
        \cos{\theta}\,
        \left(1 + \frac{M_\text{T}}{I_\text{T}}~\vec{a}\cdot\vec{b}\right)\,.
    \label{eq:pcaldef}
\end{equation}
This modulated displacement is shown schematically on the left of Fig.~\ref{fig:loopdiagram}.
The terms entering this formula are as follows: $f$ is the frequency of the power modulation, $P(f)$ is the power modulation amplitude, $c$ is the speed of light, $s(f)$ is the mechanical compliance of the suspended mirror, $\theta\simeq8.8^\circ$ is the angle of incidence on the mirror, $M_\text{T} = 39.6$\,kg and $I_\text{T} = 0.415\text{\,kg\,m}^2$ are the mass and rotational moment of inertia of the mirror, and $\vec{a}$ and $\vec{b}$ are displacement vectors from the center of the optical surface to the photon calibrator center of force and the main interferometer beam, respectively.
These displacements determine the amount of unwanted induced rotation of the mirror.

The compliance $s(f)$ of the suspended mirror can be approximated by treating the mirror as rigid body that is free to move along the optical axis of the arm cavity: $s(f) \simeq -1/[M_\text{T}(2\piup f)^2]$.
Cross-couplings between other degrees of freedom of the multi-stage suspension system, however, require that $s(f)$ be computed with a full, rigid-body model of the quadruple suspension. 
This model has been validated by previous measurements~\cite{P1200182} and is assumed to have negligible uncertainty.

Significant sources of photon calibrator uncertainty include the NIST calibration of the reference standard (0.5\%), self-induced test mass rotation uncertainty (0.4\%), and uncertainty of the optical losses along the projection and reflection paths (0.4\%).
The overall $1\sigma$ uncertainty in the displacement induced by the photon calibrator, $x_{\text{T}}^{\text{(PC)}}(f)$, is ${\simeq}\,0.8\%$. 

\section{Actuation function calibration} 
\label{sec:ActuatorCalibration}

\begin{figure}[t]
    \centering
    \includegraphics[width=\columnwidth]{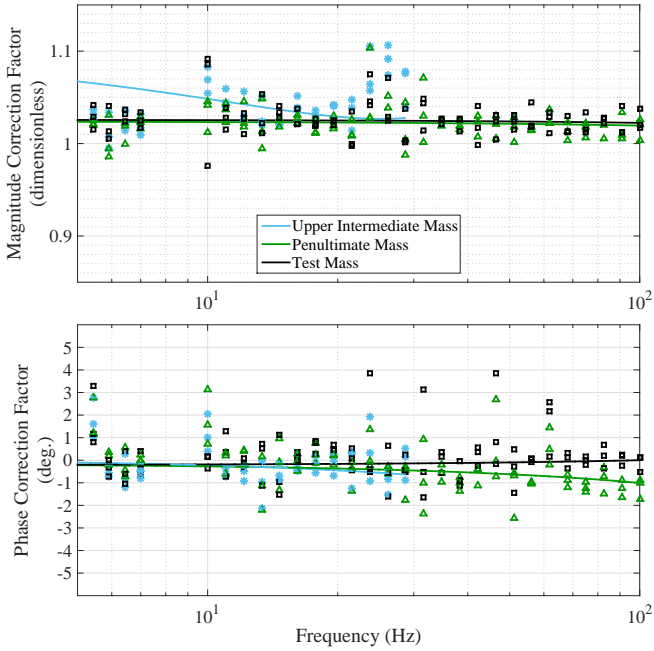}
    \caption{\label{fig:H1L1PCAL}Measured frequency-dependent correction factors, $\zeta_{i}^{\text{(fd)}}$, for the actuators of the lower three stages of the H1 suspension (symbols) and corresponding fits, $\zeta_{i}^{\text{(fd,fit)}}$ (solid lines).
    Only data up to 100~Hz for the bottom two stages were collected because the sensing function dominates the actuation function above $\sim$45\,Hz. Data for the upper intermediate mass is presented only up to 30~Hz because the actuation function for this stage is attenuated sharply above $\sim$5\,Hz.}
\end{figure}

\begin{figure}[t]
    \centering
    \includegraphics[width=\columnwidth]{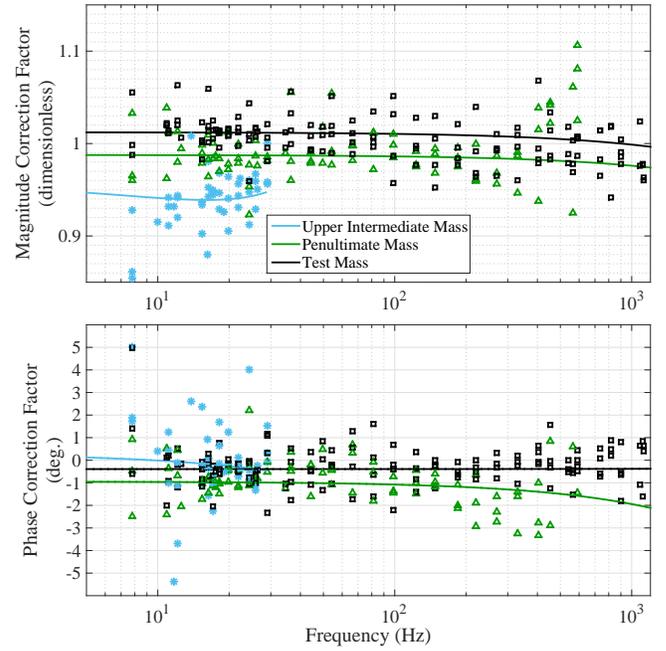}
    \caption{\label{fig:L1L1PCAL}Measured frequency-dependent correction factors, $\zeta_{i}^{\text{(fd)}}$, for the actuators of the lower three stages of the L1 suspension (symbols) and corresponding fits, $\zeta_{i}^{\text{(fd,fit)}}$ (solid lines).
    Data collected up to 1.2\,kHz confirms the expected frequency dependence of the correction factors for the bottom two stages. Data for the upper intermediate mass is presented up to 30\,Hz because the actuation function for this stage is attenuated sharply above $\sim$5\,Hz.}
\end{figure}

The actuation strength for the $i$th suspension stage, $[\mathcal{K}_i A_{i}(f)]^\text{(meas)}$, can be determined by comparing the interferometer's response, $d_\text{err}(f)$, to an excitation from that suspension stage's actuator, $\text{exc}_i(f)$, with one from the photon calibrator, $x_\text{T}^\text{(PC)}(f)$,
\begin{equation}
    [\mathcal{K}_i A_{i}(f)]^\text{(meas)} = \frac{x_\text{T}^\text{(PC)}(f)}{d_\text{err}(f)} \times \frac{d_\text{err}(f)}{\text{exc}_{i}(f)}\,.
    \label{eq:pcalmeas}
\end{equation}
Figs.~\ref{fig:H1L1PCAL} and~\ref{fig:L1L1PCAL} show the collection of these measurements for the H1 and L1 interferometers in the form of correction factors, $\zeta_{i}^\text{(fd)} =  [\mathcal{K}_i A_{i}(f)]^\text{(meas)} /  [\mathcal{K}_i A_{i}(f)]^\text{(model)}$. The collection includes the reference measurement and subsequent measurements normalized by any scalar, time-dependent correction factors, $\zeta_{i}^{\text{(td)}}$. These data are used to create the fit, $\zeta_{i}^\text{(fd,fit)}$, and estimate the actuation components of the statistical uncertainty $\sigma_{q_{i}~q_{j}}$.

As described in Sec.~\ref{sec:model}, the actuation function, and therefore its uncertainty and error, only contribute significantly to the uncertainty estimate for $h$ below $\sim$45\,Hz, which is the unity gain frequency for the differential arm length servo.
While there are no data at frequencies above 100\,Hz for H1, the L1 high-frequency data confirm that above 100\,Hz, frequency-dependent deviations from the model are small.

There are larger frequency-dependent errors in the models for the upper intermediate stages $\mathcal{K}_{\text{U}}A_{\text{U}}$ for both detectors.
Additional measurements, not explicitly included in this paper, have shown that these result from unmodeled mechanical resonances as well as the non-negligible inductance of the electromagnetic coil actuators.
As shown in Fig.~\ref{fig:ActuationModel}, however, the actuation strength of the upper intermediate mass is attenuated sharply above $\sim$5~Hz by $F_{\text{U}}$.
It therefore does not substantially impact the overall actuation model in the relevant GW frequency band.

A systematic photon calibrator error would result in an overall error in the calibrated detector strain output.
To investigate the possibility of such unknown systematic errors, two alternative calibration methods were employed.
This is similar to what was done during initial LIGO~\cite{P0900155}.
One alternative method uses a radio-frequency oscillator reference and 532~nm laser light resonating in the interferometer arm cavities to calibrate the suspension actuators.
The other method, which was also used during initial LIGO, uses the wavelength of the 1064~nm main laser light as a length reference.
Their comparison with the photon calibrator is discussed in Appendix~\ref{app:ActuatorChecks}.
No large systematic errors were identified, but the accuracy of the alternate measurements is currently limited to $\sim$10\%.

\svnid{$Id: SensorCalibration.tex 2731 2016-02-18 19:01:53Z evan.goetz@LIGO.ORG $}

\section{Sensing Function calibration}
\label{sec:SensorCalibration}

The sensing function, $C^\text{(meas)}(f)$, can be measured directly by compensating the interferometer response to photon calibrator displacement, $d_{\text{err}}(f) / x_{\text{T}}^{(\text{PC})}(f)$, for the differential arm length control suppression, $[1 + G(f)]$,
\begin{equation}
    C^\text{(meas)}(f)
        = \big[1 + G(f)\big] \times \frac{d_\text{err}(f)}{x_\text{T}^\text{(PC)}(f)}\, , \label{eq:sensing}
\end{equation}
where $G(f)$ is measured independently with the calibrated actuator. 

\begin{figure}[t]
    \centering
    \includegraphics[width=\columnwidth]{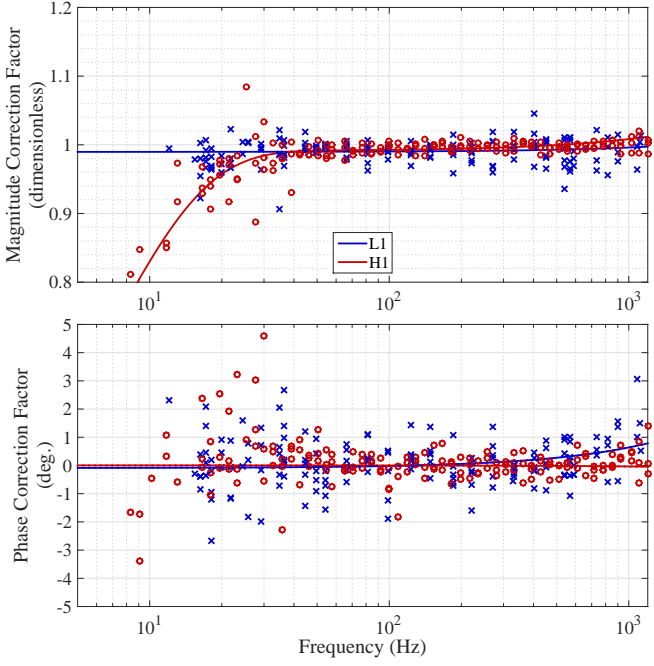}
    \caption{Measured frequency-dependent sensing function correction factors, $\zeta_{i}^{\text{(fd)}}$, for L1 (blue crosses) and H1 (red circles) and their fits, $\zeta_{i}^{\text{(fd,fit)}}$.\label{fig:SensingResiduals}}
\end{figure}

Figure~\ref{fig:SensingResiduals} shows the collection of these measurements for H1 and L1 in the form of correction factors, $\zeta_{C}^{\text{(fd)}} = C^\text{(meas)}(f) / C^\text{(model)}(f)$, appropriately normalized with time-dependent correction factors, $\zeta_{i}^{\text{(td)}}$. Corresponding fits to the frequency-dependent correction factors, $\zeta_{C}^{\text{(fd,fit)}}$, are also shown. Together, these are used to establish the sensing components of the statistical uncertainty, $\sigma_{q_{i}~q_{j}}$. 

The frequency-dependent correction factor seen in H1 exposes detuning of its signal recycling cavity~\cite{P1500277}, resulting from undesired optical losses. 
Such detuning modifies the interferometric response but is not included in the sensing function model (Eq. \ref{eq:sensing}). 
The sensing function contribution to the response function, $R(f)$, only dominates above the unity gain frequency of the differential arm length servo ($f > 45\,\text{Hz}$). 
As such, this correction factor becomes negligible when folded into the overall systematic error.

\svnid{$Id: DynamicUncertainty.tex 2703 2016-02-11 04:39:22Z evan.hall@LIGO.ORG $}

\section{Time-Dependent Systematic Errors}
\label{sec:dynamic}
The scalar calibration parameters $\mathcal{K}_{\text{C}}, f_{\text{C}}$, and $\mathcal{K}_{\text{T}}$ have been found to vary slowly as a function of time~\cite{P1600063}.
Changes in these parameters are continuously monitored from the calibration lines observed in $d_\text{err}$; these lines are injected via the photon calibrator and suspension system actuators.
The amplitude of each calibration line is tuned to have a signal-to-noise ratio (SNR) of $\sim$100 for a ten-second Fourier transform of $d_{\text{err}}$.
The calibration lines are demodulated, and their complex ratios are stored at a rate of 16\,Hz.
Running means of the complex ratios are computed over 128\,s of this data, and are used to compute the scalar parameter as a function of time.
The length of the running mean was chosen to reduce statistical uncertainty while still maintaining signal integrity for the chosen amplitudes, and to reduce the effect of non-Gaussian noise transients in the interferometer.

The optical parameters $\mathcal{K}_{\text{C}}$ and $f_{\text{C}}$ change in response to variations in the alignment or the thermal state of the interferometer optics. 
The most dramatic changes occur over the course of the few minutes immediately after the interferometer achieves resonance, when the interferometer's angular control system is settling and the optics are coming into thermal equilibrium.

Variations in $\mathcal{K}_\text{T}$ occur due to the slow accumulation of stray ions onto the fused silica test mass~\cite{P1100208, hewitson2007charge}.
Test mass charging thus creates a slow change in the actuation gain, which takes several days to cause an observable change.
The upper stage actuation gains, $\mathcal{K}_{\text{P}}$ and $\mathcal{K}_{\text{U}}$, are also monitored, but the measurements do not show time-dependent variations that are larger than the precision of the tracking measurements.

Changes in the gains $\mathcal{K}_{i}$ are represented by time-dependent correction factors, $\kappa_{i}(t) = 1 + \delta \mathcal{K}_{i}(t) / \mathcal{K}_{i} \in \zeta_{i}^{\text{(td)}}$. 
Changes in the pole frequency, however, are reported as an absolute change: $f_{\text{C}}(t) = f_{\text{C}} + \delta f_{\text{C}}$. 
Time-dependence in $f_{\text{C}}$ results in a time-dependent, frequency-dependent correction factor $\zeta_{f_{C}}^{\text{(td)}}(f)$, determined by taking the ratio of two normalized, single-pole transfer functions, one with $f_{\text{C}}$ at the reference time and the other with $f_\text{C}$ at the time of relevant observational data.
All time-dependent correction factors also have statistical uncertainty, which is included in $\sigma_{q_{i}~q_{j}}$.

Measurements to be used as references for the interferometer models were made 3 days prior and 1 day prior to GW150914 at H1 and L1, respectively.
Since the charge accumulation on the test mass actuators is slow, any charge-induced changes in the test mass actuation function parameters during these few days was less than 1\%.
At the time of GW150914, H1 had been observing for 2 hours and L1 had been observing for 48 minutes, so both detectors had achieved stable alignment and thermal conditions.
We thus expect that sensing function errors were also very small, though they fluctuate by a few percent around the mean value during normal operation.
This level of variation is consistent with the variation measured during the {\OBSSTART} to {\OBSEND} observation period.
The correction factors measured at the time of GW150914 are shown in Table~\ref{tbl:kappas}.

\begin{table}
    \caption{\label{tbl:kappas}Dimensionless correction factors $\kappa_{i}$ and systematic error in cavity pole frequency, and their associated statistical uncertainties (in parenthesis) during GW150914.}
    \begin{ruledtabular}
    \begin{tabular}{l D{.}{.}{1,3} D{.}{.}{1,1} D{.}{.}{1,3} D{.}{.}{1,1}}
            & \multicolumn{2}{c}{\textbf{H1}} & \multicolumn{2}{c}{\textbf{L1}} \\
            \cline{2-3}
            \cline{4-5}
            & \multicolumn{1}{c}{Mag.} & \multicolumn{1}{c}{Phase (deg.)} & \multicolumn{1}{c}{Mag.} & \multicolumn{1}{c}{Phase (deg.)} \\
        \hline
        $\kappa_\text{T}$  & 1.041(2)
                                          & -0.7(1)
                                          & 1.012(2)
                                          & -1.2(1)    \\
        $\kappa_\text{PU}$  & 1.022(2)
                                          & -1.3(2)
                                          & 1.005(3)
                                          & -1.5(2)    \\
        $\kappa_\text{C}$   & 1.001(3)
                                          & \text{N/A}
                                          & 1.007(3)
                                          & \text{N/A} \\
        $\delta f_\text{C}$ (Hz)          & -8.1(1.4)
                                          & \text{N/A}
                                          & 0.5(1.9)
                                          & \text{N/A}            
    \end{tabular}
    \end{ruledtabular}
\end{table}

\svnid{$Id: TotalUncertainty.tex 2730 2016-02-18 19:00:49Z jeffrey.kissel@LIGO.ORG $}

\section{Estimate of Total Uncertainty}
\label{sec:total}

The statistical uncertainty of all model parameters are combined to form the total statistical uncertainty of the response function,
\begin{equation}
    \sigma_{R}^{2}(f) = \sum_{q_i}\sum_{q_j}\left(\frac{\partial R(f)}{\partial q_i}\right)
        \left(\frac{\partial R(f)}{\partial q_j}\right)\sigma_{q_i\,q_j}\,,
\end{equation}
where $\partial R(f)/\partial q_i$ is the partial derivative of $R$ with respect to a given parameter $q_i$. 

The total systematic error in the response function, $\delta R$, represented as a correction factor, $1 + \delta R/R$, is evaluated by computing the ratio of the response function with its parameters evaluated with and without time- and frequency-dependent actuation and sensing correction factors
\begin{equation}
    1 + \frac{\delta R(f,t)}{R(f)}
        = \frac{R(f\,; q_{1}, q_{2}, \ldots, q_{n})}{R(f,t\,; q_{1} + \delta q_{1}, q_{2} + \delta q_{2}, \ldots,q_{n} + \delta q_{n})}\,.
\end{equation}
Therefore, the response function correction factor quantifies the systematic error of the calibrated detector strain output at the time of GW150914.

Measurements made during and after the observation period revealed that the estimate of $x_\text{T}^\text{(PC)}$ also includes systematic errors $\delta x_\text{T}^\text{(PC)}$, resulting in frequency-independent correction factors of 1.013 and 1.002 for H1 and L1, respectively.
These errors affect both the actuation and sensing function, and are included accordingly with other known systematic errors in the response function.

\begin{figure}[t]
    \centering
    \includegraphics[width=\columnwidth]{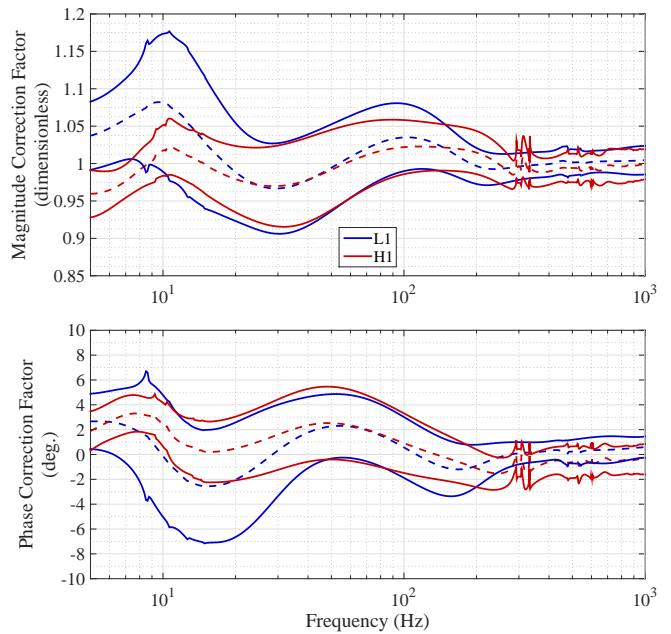}
    \caption{Known systematic error and uncertainty for the response function $R(f)$ at the time of GW150914, expressed as a complex correction factor $1+\delta R(f, t)/R(f)$ (dashed lines) with surrounding uncertainty $\pm~\sigma_{R}(f)$ (solid lines).
        The upper panel shows the magnitude, and the lower panel shows the phase. 
        The solid lines define the 68\% confidence interval of the precision and accuracy of our estimate of $h(t)$. \label{fig:total}}
\end{figure}

Figure~\ref{fig:total} shows the total statistical uncertainty and correction factors for each interferometer's response function, $R(f)$, at the time of GW150914 and defines the 68\% confidence interval on the accuracy and precision of $h(t)$. 
Systematic errors at low frequency are dominated by the systematic errors in the actuation function, whereas at high frequencies, the systematic error is dominated by the sensing function systematic error.
The frequency dependence of the sensing and actuation models, and of the uncertainties presented here, is expected to be smoothly varying in the 20\,Hz to 1\,kHz band.
For all frequencies relevant to GW150914, between 20\,Hz and 1\,kHz, the uncertainty is less than \hl{10\%} in magnitude and \hl{$10^\circ$} in phase.
The comparison of measurements with models presented in Sec.~\ref{sec:ActuatorCalibration} and Sec.~\ref{sec:SensorCalibration} of this paper are consistent with that expectation.

\svnid{$Id: appA_ScaleDoubleChecks.tex 2402 2016-01-27 19:35:57Z jeffrey.kissel@LIGO.ORG $}

\section{Inter-site Timing Accuracy}
\label{sec:Timing}

Digital signals $d_{\text{err}}$ and $d_{\text{ctrl}}$ are derived from signals captured by analog-to-digital converters as a part of the LIGO data acquisition system~\cite{bork2011new} and are stored in a mass data storage system which records these signals for later analysis.
The LIGO timing system~\cite{P0900265} provides the reference timing information for the data acquisition system, which records the data with an associated Global Positioning System (GPS) time stamp.

Each detector's timing system uses a single Trimble Thunderbolt E GPS receiver as the timing reference. 
Additional GPS receivers and one cesium atomic clock serve as witness clocks independently monitoring the functionality of the main GPS reference. 
Once a second, timing comparators monitor the clock edge differences (modulo one second) between the main GPS receiver and the witness clocks with sub-microsecond accuracy. 
We did not observe any anomaly at the time of GW150914.

Large absolute timing offsets must also be ruled out with the GPS units at each site, which may be out of range of the timing comparators. 
The GPS units produce IRIG-B time code signals which can be recorded by the data acquisition system. 
The IRIG-B time code provides a map from the acquisition system's GPS time to Coordinated Universal Time (UTC). 
At the time of GW150914, IRIG-B signals generated by the witness GPS receivers were recorded at H1. 
At L1, IRIG-B signals generated by the reference GPS receiver were recorded as a self-consistency check. 
Throughout all 38~days of observation, no large offset was observed between any witness or reference IRIG-B signals and UTC at either site.
Witness receivers were added at L1 after the initial 38 days, and their IRIG-B codes showed no inconsistency. 
We expect the uncertainty in this comparison to be smaller than the $1\,\muup\text{s}$ specifications of typical GPS systems ~\cite{2007GPSperformance, 2008GPSSPS, GPS}.

Additional monitoring is performed to measure any potential timing offset which may occur internally between the timing system and the analog-to-digital and digital-to-analog converters. 
This monitoring system is described in detail in \cite{P0900265}, but summarized here.
Two analog, sinusoidal diagnostic signals at 960 and 961 Hz are generated by each data acquisition unit. 
The beat note of these two sine waves and all ADCs and DACs in the unit itself are synchronized with a one-pulse-per-second signal sent from the reference GPS receiver via optical fiber with accuracy at the micro-second level. 
Within a given converter, the channel-to-channel synchronization is well below this uncertainty \cite{GeneralStandardsADC, GeneralStandardsDAC}.
The known diagnostic waveform is also injected into a subset of analog-to-digital converters in each data acquisition unit. 
The recorded waveform can then be compared against the acquisition time stamp, accounting for the expected delay. 
Any discrepancy would reveal that data acquisition unit's timing is offset relative to the timing reference.
The diagnostic signals on units directly related to the estimated detector strain $h(t)$---the GW readout and photon calibrator photodetectors---are recorded permanently. 
These signals were examined over a 10-minute window centered on the time of GW150914. 
In both detectors, these offsets were between 0.6 and $0.7\,\muup\text{s}$ depending on the unit, with the standard deviation smaller than 1\,ns in each given unit.
Although potential timing offsets between different channels on the same analog-to-digital-converter board were not measured, there is no reason to believe that there were any timing offsets larger than a few microseconds.

Based on these observations we conclude that the LIGO timing systems at both sites were working as designed and internally consistent over all 38 days of observation.
Even if the most conservative estimate is used as a measure of caution, the absolute timing discrepancy from UTC, and therefore between detectors, was no larger than $10\,\muup\text{s}$.
The impact of this level of timing uncertainty is discussed in Section \ref{sec:discussion}.

\svnid{$Id: Conclusion.tex 3009 2016-06-16 16:49:10Z jeffrey.kissel@LIGO.ORG $}

\section{Impact of calibration uncertainties on GW150914} 
\label{sec:discussion}

The total uncertainty in $h(t)$ reported in Section \ref{sec:total} is less than 10\% in magnitude 
and 10$^\circ$ in phase from $20\,\text{Hz}$ to $1\,\text{kHz}$ for the entire 38 calendar days of observational data 
during which GW150914 was observed. The astrophysical searches used for detecting events like GW150914 
are not limited by this level of calibration uncertainty~\cite{GW150914-BURST, GW150914-CBC}.

Calibration uncertainties directly affect the estimation of the source parameters associated with events like GW150914.
The amplitude of the gravitational wave depends on both the luminosity distance and the orbital inclination of the source, so uncertainty in the magnitude of the calibration, determined by the photon calibrator, directly affects the estimation of the luminosity distance.
The luminosity distance also depends strongly, however, on the orbital inclination of the binary source, 
which is poorly constrained by the two nearly co-aligned Advanced LIGO detectors. 
Thus, the 10\% uncertainty in magnitude does not significantly degrade 
the accuracy of the luminosity distance for GW150914~\cite{GW150914-PARAMESTIM}. 
The absolute scale is cross-checked with two additional calibration methods, 
one referenced to the main laser wavelength and another 
referenced to a radio-frequency oscillator (Appendix~\ref{app:ActuatorChecks}). 
Each method is able to confirm the scale at the 10\% level in both detectors, 
comparable to the estimate of total uncertainty in absolute scale.

An uncertainty of 10\% in the absolute strain calibration results in a $\sim$30\% uncertainty 
on the inference of coalescence rate for similar astrophysical systems~\cite{GW150914-RATES}. 
Since the counting uncertainty inherent in the rate estimation surrounding GW150914 is larger than the 30\% uncertainty in rates induced by the calibration uncertainty, the latter does not yet limit the rate estimate.

Estimating the sky-location parameters depends partially on the inter-site accuracy of the detectors' timing systems~\cite{P0900090}. 
These systems, and the consistency checks that were performed on data containing GW150914, 
are described briefly in Section~\ref{sec:SensorCalibration}. 
The absolute time of detectors' data streams is accurate to within $10\,\muup\text{s}$, which does not limit 
the uncertainty in sky-location parameters for GW150914~\cite{GW150914-PARAMESTIM,GW150914-EMFOLLOW}. 
Further, the phase uncertainty of the response function as shown in Section~\ref{sec:total} is much larger than the corresponding phase uncertainty arising from intra-site timing in the detection band (a $\pm 10\,\muup\text{s}$ timing uncertainty corresponds to a phase uncertainty of 0.36$^{\circ}$ at $100\,\text{Hz}$).

All other astrophysical parameters rely on the accuracy of each detector's output calibration as a function of frequency.
The physical model of the frequency dependence underlying this uncertainty was not directly available to the parameter estimation procedure
at the time of detection and analysis of GW150914. 
Instead, a preliminary model of the uncertainty's frequency dependence was used, the output of which was a smooth, 
parameterized shape over the detection band~\cite{T1400682,GW150914-PARAMESTIM}. 
The parameters of the preliminary model were given Gaussian prior distributions such that its output was consistent 
with the uncertainties described in this paper.
Comparison between the preliminary model and the physical model presented in this paper have shown that the preliminary model  
is sufficiently representative of the frequency dependence. 
In addition, its uncertainty has been shown not to limit the estimation of astrophysical parameters for GW150914~\cite{GW150914-PARAMESTIM}.

\section{Summary and Conclusions} 
\label{sec:conclusion}

In this paper, we have described how the calibrated strain estimate $h(t)$ is produced 
from the differential arm length readout of the Advanced LIGO detectors.
The estimate is formed from models of the detectors' actuation and sensing systems 
and verified with calibrated, frequency-dependent excitations via radiation pressure actuators at reference times. 
This radiation pressure actuator relies on a NIST-traceable 
laser power standard and knowledge of the test mass suspension dynamics, which are both known at the 1\% level.
The reference and subsequent confirmation measurements inform the static, 
frequency-dependent systematic error and statistical uncertainty in the estimate of $h(t)$.
Time-dependent correction factors to certain model parameters are monitored 
with single-frequency excitations during the entire observation period. 
We report that the value and statistical uncertainty of these time-dependent factors 
are small enough that they do not impact astrophysical results throughout the period from {\OBSSTART} to {\OBSEND}, 2015.

The reference measurements and time-dependent correction factors are used to estimate 
the total uncertainty in $h(t)$, which is less than 10\% in magnitude 
and 10$^\circ$ in phase from $20\,\text{Hz}$ to $1\,\text{kHz}$ 
for the entire 38 calendar days of observation during which GW150914 was observed. 
This level of uncertainty does not significantly limit the estimation of source parameters associated with GW150914.
We expect these uncertainties to remain valid up to 2\,kHz once the forthcoming calibration for the full LIGO observing run is complete.

Though not yet the dominant source of error, based on the expected sensitivity improvement of Advanced LIGO~\cite{P1200087}, calibration uncertainties may limit astrophysical measurements in future observing runs. In the coming era of numerous detections of gravitational waves from diverse sources, accurate estimation of source populations and properties will depend critically on the accuracy of the calibrated detector outputs of the advanced detector network. In the future, the calibration physical model and its uncertainty will be directly employed in the astrophysical parameter estimation procedure, which will reduce the impact of this uncertainty on the estimation of source parameters. We will continue to improve on the calibration accuracy and precision reported here, with the goal of ensuring that future astrophysical results are not limited by calibration uncertainties as the detector sensitivity improves and new sources are observed.

\begin{acknowledgments}
The authors gratefully acknowledge the support of the United States
National Science Foundation (NSF) for the construction and operation of the
LIGO Laboratory and Advanced LIGO as well as the Science and Technology Facilities Council (STFC) of the
United Kingdom, the Max-Planck-Society (MPS), and the State of
Niedersachsen/Germany for support of the construction of Advanced LIGO 
and construction and operation of the GEO600 detector. 
Additional support for Advanced LIGO was provided by the Australian Research Council.
The authors gratefully acknowledge the Italian Istituto Nazionale di Fisica Nucleare (INFN),  
the French Centre National de la Recherche Scientifique (CNRS) and
the Foundation for Fundamental Research on Matter supported by the Netherlands Organisation for Scientific Research, 
for the construction and operation of the Virgo detector
and the creation and support  of the EGO consortium. 
The authors also gratefully acknowledge research support from these agencies as well as by 
the Council of Scientific and Industrial Research of India, 
Department of Science and Technology, India,
Science \& Engineering Research Board (SERB), India,
Ministry of Human Resource Development, India,
the Spanish Ministerio de Econom\'ia y Competitividad,
the Conselleria d'Economia i Competitivitat and Conselleria d'Educaci\'o, Cultura i Universitats of the Govern de les Illes Balears,
the National Science Centre of Poland,
the European Union,
the Royal Society, 
the Scottish Funding Council, 
the Scottish Universities Physics Alliance, 
the Lyon Institute of Origins (LIO),
the National Research Foundation of Korea,
Industry Canada and the Province of Ontario through the Ministry of Economic Development and Innovation, 
the National Science and Engineering Research Council Canada,
the Brazilian Ministry of Science, Technology, and Innovation,
the Research Corporation, 
Ministry of Science and Technology (MOST), Taiwan
and
the Kavli Foundation.
The authors gratefully acknowledge the support of the NSF, STFC, MPS, INFN, CNRS and the
State of Niedersachsen/Germany for provision of computational resources.
This article has been assigned the LIGO document number P1500248.
\end{acknowledgments}

\appendix

\svnid{$Id: CrossChecks.tex 2730 2016-02-18 19:00:49Z jeffrey.kissel@LIGO.ORG $}

\section{Photon calibrator cross-check}
\label{app:ActuatorChecks}

It is essential to rule out large systematic errors in the photon calibrator by comparing it against fundamentally different calibration methods.
For Advanced LIGO, two alternative methods have been implemented.
One is based on a radio-frequency oscillator and the other based on the laser wavelength. Each of them is described below.

\subsection{Calibration via radio-frequency oscillator}
As part of the control sequence to bring the interferometer to resonance, the differential arm length is measured and controlled using two auxiliary green lasers with a wavelength of 532\,nm~\cite{P1400177,P1400105,Mullavey2012OE}.
Although designed as part of the interferometer controls, this system can provide an independent measure of the differential arm length.

The two green lasers are offset from each other in frequency by 158\,MHz.
The frequency of each is independently locked to one of the arm cavities with a control bandwidth of several kilohertz.
Therefore, the frequency fluctuations of each green laser are proportional to the length fluctuations of the corresponding arm cavity through the relation $\Delta \nu_\text{g}/\nu_\text{g} \approx \Delta L/L$, where $\nu_\text{g}$ is the frequency of either of the auxiliary lasers~\cite{Izumi2012JOSAA}.
Beams from these two lasers are interfered and measured on a photodetector, producing a beat-note close to 158\,MHz.
As the differential arm length varies, the beat-note frequency shifts by the amount defined by the above relation.
This shift in the beat-note frequency is converted to voltage by a frequency discriminator based on a voltage controlled oscillator at a radio frequency.
Therefore the differential arm length can be calibrated into physical displacement by calibrating the response of the frequency discriminator.

A complicating factor with this method is the limited availability.
This method is only practical for calibration in a high noise interferometer configuration because sensing noise is too high.
Another set of measurements is thus required to relate the high noise actuators to the ones configured for low noise observation.
These extra measurements are conducted in low noise interferometer state where both high and low noise actuators are excited. 
Since both excitations are identically suppressed by the control system, simply comparing their responses using the readout signal $d_\text{err}$ allows for propagation of the calibration.
In summary, one can provide an independent calibration of every stage of the low noise actuator by three sets of measurements:
\begin{equation}
    [\mathcal{K}_{i}A_{i}(f)]^\text{(rf)} =  \left(\frac{\Delta L}{\text{exc}_\text{HR}(f)} \right) \times \left(\frac{\text{exc}_\text{HR}(f)}{d_\text{err}(f)} \right)\times \left( \frac{d_\text{err}(f)}{\text{exc}_{i}(f)} \right),
\end{equation} 
where $\text{exc}_\text{HR}$ is digital counts applied to excite a high noise actuator.
The first term on the right hand side represents the absolute calibration of the high noise actuator, and the final two ratios represent the propagation of the calibration in low noise interferometer state.

\begin{figure}
    \centering
    \includegraphics[width=\columnwidth]{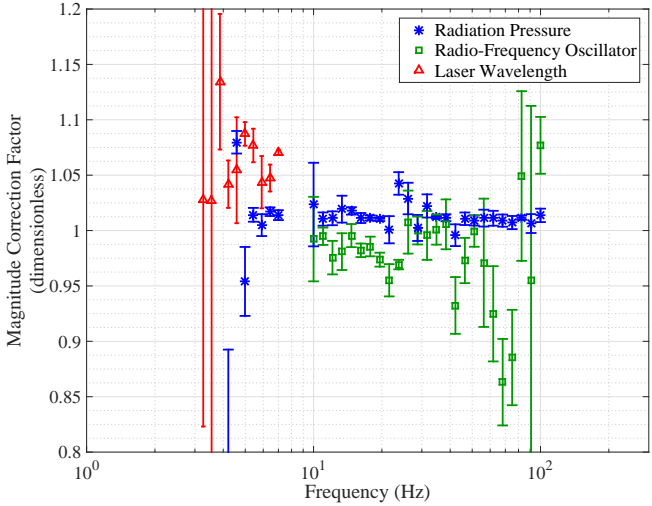}
    \caption{Comparison between radiation pressure, radio frequency oscillator, and laser wavelength calibration techniques, displayed as $[\mathcal{K}_{\text{T}}A_{\text{T}}(f)]^{(\text{method})} / [\mathcal{K}_{\text{T}}A_{\text{T}(f)}]^{\text{(model)}}$, for the test mass stage of the H1 interferometer. Only statistical uncertainty is shown; systematic errors for individual methods are not shown. \label{fig:H1EYL3ScaleFactor}}
\end{figure}

\begin{figure}
    \centering
    \includegraphics[width=\columnwidth]{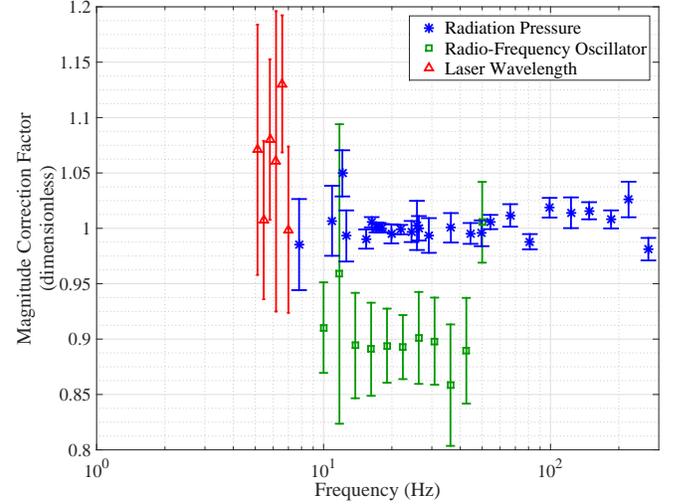}
    \caption{Comparison between radiation pressure, radio frequency oscillator, and laser wavelength calibration techniques, displayed as $[\mathcal{K}_{\text{T}}A_{\text{T}}(f)]^{(\text{method})} / [\mathcal{K}_{\text{T}}A_{\text{T}(f)}]^{\text{(model)}}$, for the test mass stage of the L1 interferometer. Only statistical uncertainty is shown; systematic errors for individual methods are not shown. \label{fig:L1EYL3ScaleFactor}}
\end{figure}

\subsection{Calibration via laser wavelength}
The suspension actuators can be calibrated against the main laser wavelength ($\lambda_\text{r} = 1064\,\text{nm}$) using a series of different optical topologies.
The procedure is essentially the same as the procedure for initial gravitational wave detectors~\cite{accadia2011calibration, P0900120}.

First, the input test masses and the beamsplitter are used to form a simple Michelson topology, which allows the input test mass suspension actuators to be calibrated against the main laser wavelength.
Then, a laser (either main or auxiliary green) is locked to the Fabry--P\'{e}rot cavity formed by the X-arm input and end test masses.
This allows the end test mass actuators to be calibrated against the corresponding input test mass actuators.
Finally, in the full optical configuration, the low noise suspension actuators (of the Y-arm end test mass) are calibrated against the X-arm end test mass suspension actuators.

In Advanced LIGO, one practical drawback is the narrow frequency range in which this technique is applicable.
Not all input test masses suspensions have actuation on the final stage, so the the penultimate mass suspension actuators must be used instead. 
This limits the frequency range over which one can drive above the displacement sensitivity of the Michelson.
The penultimate stage actuators themselves are also weak, further reducing the possible signal-to-noise ratio of the fundamental measurement.
As a consequence, the useable frequency range is limited to below 10\,Hz.

\subsection{Results and discussion}

Figures~\ref{fig:H1EYL3ScaleFactor} and~\ref{fig:L1EYL3ScaleFactor} show the correction factor for $\mathcal{K}_{T} A_{T}$.
Only the test mass stage is shown for brevity.
This comparison was done for all three masses of actuation system and show similar results.
With the correction factors of both independent methods (radio frequency oscillator and laser wavelength) within 10\% agreement with that as estimated by radiation pressure (again, for all stages of actuation), we consider the absolute calibration of the primary method confirmed to that 10\% level of accuracy.
At this point, the independent methods are used merely to bound the systematic error on the radiation pressure technique's absolute calibration; considerably less effort and time were put into ensuring that all discrepancies and systematic errors within the independent method were well-quantified and understood. 
Only statistical uncertainty---based on coherence for each compound-measurement point in each method---is shown, because the systematic error for these independent methods have not yet been identified or well-quantified.
Refinement and further description of these techniques is left for future work.

\bibliographystyle{apsrev4-1}
\bibliography{P1500248_EarlyaLIGOCalUncertainty,GW150914/macros/GW150914_refs}

\end{document}